\documentclass[english,APS, A4, Engish,superscriptaddress, twocolumn]{revtex4-1}
\usepackage[T1]{fontenc}
\usepackage[latin9]{inputenc}
\usepackage{color}
\usepackage{babel}
\usepackage{amstext}
\usepackage{amssymb}
\usepackage{graphicx}
\usepackage{esint}
\usepackage[unicode=true,pdfusetitle,
 bookmarks=true,bookmarksnumbered=false,bookmarksopen=false,
 breaklinks=true,pdfborder={0 0 0},pdfborderstyle={},backref=false,colorlinks=true]
 {hyperref}
\hypersetup{
 linkcolor=blue, citecolor=cyan}

\makeatletter

\providecommand{\tabularnewline}{\\}

\usepackage{xcolor}
\newcommand{\trento}{T\raisebox{-0.5ex}{R}ENTo}

\makeatother

\begin{document}
\title{Spin polarization of $\Lambda$ hyperons along beam direction
in p+Pb collisions at $\sqrt{s_{NN}}=8.16$ TeV using hydrodynamic approaches}
\author{Cong Yi}
\email{congyi@mail.ustc.edu.cn}

\affiliation{Department of Modern Physics, University of Science and Technology
of China, Hefei, Anhui 230026, China}
\author{Xiang-Yu Wu}
\email{xiangyu.wu2@mail.mcgill.ca}

\affiliation{Department of Physics, McGill University, Montreal, QC, Canada H3A
2T8}
\affiliation{Department of Physics and Astronomy, Wayne State University, Detroit
MI 48201}
\author{Jie Zhu}
\affiliation{Institute of Particle Physics and Key Laboratory of Quark and Lepton
Physics (MOE), Central China Normal University, Wuhan 430079, China}
\affiliation{Fakult\"at f\"ur Physik, Universit\"at Bielefeld, D-33615 Bielefeld,
Germany}
\author{Shi Pu}
\email{shipu@ustc.edu.cn}

\affiliation{Department of Modern Physics, University of Science and Technology
of China, Hefei, Anhui 230026, China}
\affiliation{Southern Center for Nuclear-Science Theory (SCNT), Institute of Modern
Physics, Chinese Academy of Sciences, Huizhou 516000, Guangdong Province,
China}
\author{Guang-You Qin}
\email{guangyou.qin@ccnu.edu.cn}

\affiliation{Institute of Particle Physics and Key Laboratory of Quark and Lepton
Physics (MOE), Central China Normal University, Wuhan 430079, China}

\begin{abstract}
We have implemented the 3+1 dimensional CLVisc hydrodynamics model with \trento-3D initial conditions to investigate the spin polarization of $\Lambda$ hyperons along the beam direction in p+Pb collisions at $\sqrt{s_{NN}} = 8.16$ TeV. Following our previous theoretical framework based on quantum kinetic theory, we consider three different scenarios: $\Lambda$ equilibrium, $s$ quark equilibrium, and iso-thermal equilibrium scenarios. We have computed the second Fourier sine coefficients of spin polarization along the beam direction, denoted as $\left\langle P_{z} \sin 2(\phi_{p} - \Psi_{2}) \right\rangle$, with $\phi_{p} - \Psi_{2}$ being the azimuthal angle relative to the second-order event plane $\Psi_{2}$, as functions of multiplicity, transverse momentum and pseudo-rapidity in the three scenarios. Additionally, we have also computed the spin polarization along the beam direction, $P_{z}$, as a function of the azimuthal angle. We find that the spin polarization induced by thermal vorticity always provides an opposite contribution compared to the shear-induced polarization in p+Pb collisions. The total spin polarization computed by the current hydrodynamic model disagrees with the data measured by LHC-CMS experiments. Our findings imply that other non-flow effects may play a crucial role in p+Pb collisions.
\end{abstract}
\maketitle

\section{Introduction}

Relativistic nucleus-nucleus collisions generate a significant initial orbital angular momentum, which induces vortical fields perpendicular to the reaction plane in the quark-gluon plasma (QGP) and leads to the global spin polarization of $\Lambda$ and $\overline{\Lambda}$ hyperons \citep{STAR:2017ckg} and spin alignment of vector mesons \citep{STAR:2022fan} through spin-orbital coupling \citep{Liang:2004ph,Gao:2007bc}. The global spin polarization of $\Lambda$ and $\overline{\Lambda}$ hyperons can be well understood by the thermal vorticity combined with the modified Cooper-Frye formula, as simulated in various phenomenological models \citep{Becattini:2007nd,Becattini:2007sr,Becattini:2013fla,Fang:2016vpj,Karpenko:2016jyx,Xie:2017upb,Li:2017slc,Sun:2017xhx,Shi:2017wpk,Xia:2018tes,Shi:2019wzi,Fu:2020oxj,Lei:2021mvp,Ambrus:2020oiw,Vitiuk:2019rfv}.
Meanwhile, the local spin polarization of $\Lambda$ hyperons, which is the polarization along the beam and out-of-plane directions as a function of azimuthal angle, cannot be well explained solely by the contribution from thermal vorticity \citep{Voloshin:2017kqp,Liu:2019krs,Wu:2020yiz,Wu:2019eyi,Xia:2019fjf,Becattini:2019ntv,Li:2021jvn,Ivanov:2019ern,Guo:2021udq,Ayala:2021xrn,Deng:2021miw}. Later, shear-induced polarization was introduced to better understand the data \citep{Hidaka:2017auj,Liu:2020dxg,Becattini:2021suc,Liu:2021uhn,Fu:2021pok,Becattini:2021iol,Yi:2021ryh,Ryu:2021lnx,Florkowski:2021xvy,Buzzegoli:2022fxu,Becattini:2022zvf,Palermo:2022lvh,Wu:2022mkr,Fu:2022myl,Fu:2022oup,Palermo:2024tza}. For more detailed discussions, we refer to the recent reviews \citep{Gao:2020vbh,Hidaka:2022dmn,Becattini:2024uha} and references therein.

Physically, the polarization along the beam direction is related to the anisotropic flows $v_{n}$. For example, $v_{2}$ can induce a quadrupole vortical structure in the beam direction. This vortical structure eventually leads to the polarization of hyperons along the beam direction. The quadrupole $\sin 2(\phi_{p}-\Psi_2)$ structure of the spin polarization along the beam direction, with $\phi_{p}$ being the azimuthal angle relative to the second-order event plane $\Psi_{2}$, has been observed in various relativsitic nucleus-nucleus collisions, e.g., in RHIC-STAR \citep{Adam:2019srw,STAR:2023eck} and LHC-ALICE experiments \citep{ALICE:2021pzu}. More complicated sine modulation structures induced by triangular flow have also been observed in isobaric nucleus-nucleus collisions by RHIC-STAR experiments \citep{STAR:2023eck}. However, the polarization along the beam direction at $p_{T}>1.5$ GeV or in the low multiplicity region in relativistic nucleus-nucleus collisions remains puzzling \citep{STAR:2023eck}.


Recently, LHC-CMS experiments have measured the spin polarization  of $\Lambda$ hyperons along the beam direction in p+Pb collisions at $\sqrt{s_{NN}}=8.16$ TeV \citep{chenyan2024talk}. The second Fourier sine coefficient of the spin polarization of $\Lambda$ hyperons along the beam direction decreases as multiplicity increases. This contrasts with the trend of $v_{2}$. This finding challenges the picture that spin polarization along the beam direction in p+Pb collisions is induced by anisotropic flow.

Therefore, it is necessary to study the spin polarization of $\Lambda$ hyperons along the beam direction in p+Pb collisions using the models that are employed to understand the data in relativistic nucleus-nucleus collisions. In this work, we will investigate the spin polarization  of $\Lambda$ hyperons along the beam direction and its dependence on multiplicity, transverse momentum $p_{T}$, azimuthal angle $\phi_{p}-\Psi_{2}$ and  pseudo-rapidity $\eta$ in p+Pb collisions at $\sqrt{s_{NN}}=8.16$ TeV, by employing the 3+1D CLVisc hydrodynamics model with the \trento-3D initial condition.

The structure of this paper is as follows. In Sec. \ref{sec:Theoretical-Framework}, we briefly review the theoretical framework based on quantum kinetic theory, which has been widely used in the relativistic nucleus-nucleus system. Then, we introduce the numerical setup and parameters used in our simulation for the p+Pb collisions in Sec. \ref{sec:Numerical-framework}. We present our results in Sec. \ref{sec:Numerical-results} and summarize this work in Sec. \ref{sec:Summary}. Throughout this work, we adopt the metric $g_{\mu\nu}=\textrm{diag}\{+,-,-,-\}$ and the Levi-Civita tensor $\epsilon^{0123}=1$.


\begin{figure*}
\includegraphics[scale=0.5]{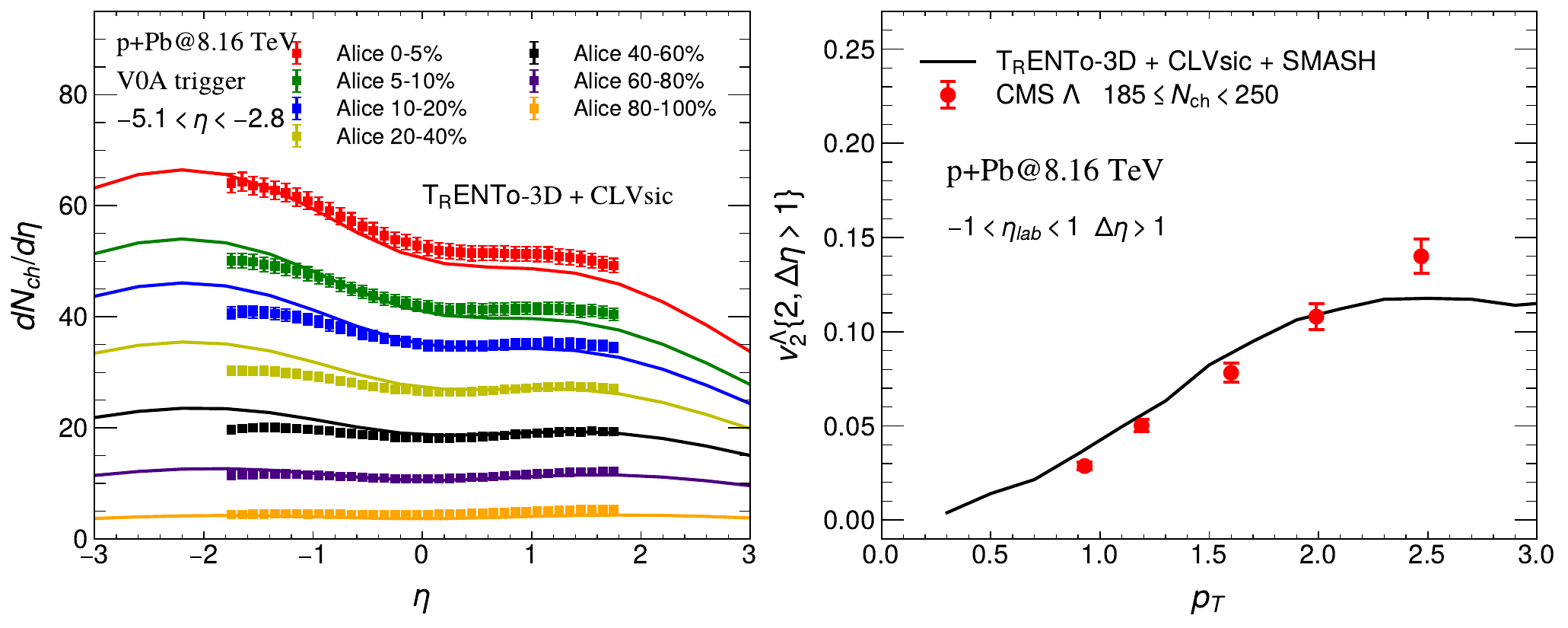}
\caption{(a) The centrality-dependent pseudo-rapidity distribution of charged hadrons and (b) the elliptic flow $v_{2}$ of $\Lambda$ hyperons in high multiplicity events as a function of $p_{T}$ in p+Pb collisions at $\sqrt{s_{NN}}=8.16$ TeV. The solid lines denote the results simulated by the 3+1D CLVisc hydrodynamics model with the \trento-3D initial condition. The experimental data are taken from Refs. \citep{CMS:2017shj,ALICE:2018wma,CMS:2018loe}. \protect\label{fig:dNdeta}}
\end{figure*}

\section{Theoretical framework \label{sec:Theoretical-Framework}}

In this section, we will briefly review the phenomenological model based on quantum kinetic theory for computing spin polarization in relativistic nucleus-nucleus collisions, following our previous works \citep{Yi:2021unq,Yi:2021ryh,Wu:2022mkr,Yi:2023tgg}. The polarization pseudo-vector for fermions $S^{\mu}(p)$ can be given by integrating the normalized axial charge current in the phase space $\mathcal{J}_{5}^{\mu}(p,X)$ over the (chemical) freezeout hyper-surface $\Sigma^{\mu}$, namely the modified Cooper-Frye formula \citep{Becattini:2013fla, Fang:2016vpj},
\begin{equation}
\mathcal{S}^{\mu}(\mathbf{p}) = \frac{\int d\Sigma \cdot p \mathcal{J}_{5}^{\mu}(p,X)}{2m \Phi(\mathbf{p})}, \label{eq:CS}
\end{equation}
where 
\begin{equation}
\Phi(\mathbf{p}) = \int d\Sigma \cdot p f_{\rm eq},
\end{equation}
where $d\Sigma^{\mu}$ is the normal vector of the (chemical) freeze-out hyper-surface elements, $p^{\mu} = (\sqrt{\mathbf{p}^{2} + m^{2}}, \mathbf{p})$ is the four-momentum of the fermions with mass $m$, and $f_{\rm eq} = \{\exp[(p^{\mu}u_{\mu} - \mu)/T] + 1\}^{-1}$ is chosen as the Fermi-Dirac distribution with $\mu$ being the (baryon) chemical potential and $T$ being the temperature. Inserting $\mathcal{J}_{5}^{\mu}$ from quantum kinetic theory \citep{Hidaka:2017auj,Yi:2021ryh} into Eq.~(\ref{eq:CS}), we find that $\mathcal{S}^{\mu}(\mathbf{p})$ can be further decomposed as
\begin{eqnarray}
\mathcal{S}^{\mu}(\mathbf{p}) & = & \mathcal{S}_{\text{thermal}}^{\mu}(\mathbf{p}) + \mathcal{S}_{\text{th-shear}}^{\mu}(\mathbf{p}), \label{eq:total}
\end{eqnarray}
where
\begin{eqnarray}
\mathcal{S}_{\text{thermal}}^{\mu}(\mathbf{p}) & = & \hbar \int d\Sigma \cdot \mathcal{N}_{p} \frac{1}{2} \epsilon^{\mu\nu\alpha\beta} p_{\nu} \varpi_{\alpha\beta}, \nonumber \\
\mathcal{S}_{\text{th-shear}}^{\mu}(\mathbf{p}) & = & \hbar \int d\Sigma \cdot \mathcal{N}_{p} \frac{\epsilon^{\mu\nu\alpha\beta} p_{\nu} n_{\beta}}{(n \cdot p)} p^{\sigma} \xi_{\sigma\alpha}, \label{eq:th_shear}
\end{eqnarray}
with
\begin{eqnarray}
\varpi_{\alpha\beta} & = & \frac{1}{2} \left[\partial_{\alpha} \left(\frac{u_{\beta}}{T}\right) - \partial_{\beta} \left(\frac{u_{\alpha}}{T}\right)\right], \nonumber \\
\xi_{\alpha\beta} & = & \frac{1}{2} \left[\partial_{\alpha} \left(\frac{u_{\beta}}{T}\right) + \partial_{\beta} \left(\frac{u_{\alpha}}{T}\right)\right],
\end{eqnarray}
being the thermal-vorticity and thermal-shear tensor, respectively, and $\mathcal{N}_{p}^{\mu} = p^{\mu} f_{\rm eq} (1 - f_{\rm eq}) / [4m \Phi(\mathbf{p})]$.
Here, the $n^{\mu}$ in $\mathcal{S}_{\text{th-shear}}^{\mu}(\mathbf{p})$ is a unit vector related to different scenarios. The decomposition in our previous works \citep{Yi:2021unq,Yi:2021ryh,Wu:2022mkr,Yi:2023tgg} is equivalent to Eq.~(\ref{eq:total}) in the absence of electromagnetic fields and by neglecting the polarization induced by $\nabla(\mu/T)$. Similar results can also be derived from linear response theory \citep{Liu:2020dxg,Liu:2021uhn,Fu:2021pok,Fu:2022myl} and quantum statistical models \citep{Becattini:2021suc,Becattini:2021iol}. 

Next, we discuss three different scenarios commonly used to study spin polarization in relativistic nucleus-nucleus collision systems.
\begin{itemize}
\item \emph{\textquotedblleft $\Lambda$ equilibrium\textquotedblright{}} scenario: In this scenario, it is assumed that $\Lambda$ hyperons reach the local (thermal) equilibrium at the freeze-out hyper-surface $\Sigma^{\mu}$ \citep{Becattini:2013fla}. Therefore, $n^{\mu}$ and $m$ in Eqs. (\ref{eq:th_shear}) are chosen as the fluid velocity $u^{\mu}$ and the mass of $\Lambda$ hyperons, respectively.
\item \emph{\textquotedblleft s quark equilibrium\textquotedblright{}} scenario: The spin 
of $\Lambda$ hyperons is assumed to be carried by the constituent $s$ quark based on the coalescence model \citep{Liang:2004ph}. Therefore, $n^{\mu}$ and $m$ in Eqs. (\ref{eq:th_shear}) are chosen as the fluid velocity $u^{\mu}$ and the mass of $s$ quarks, respectively \citep{Fu:2021pok}.
\item \emph{\textquotedblleft Iso-thermal equilibrium\textquotedblright{}} scenario: The temperature of the system at the freeze-out hyper-surface is assumed to be constant, and all terms proportional to $\nabla T$ in Eqs.~(\ref{eq:th_shear}) vanish \citep{Becattini:2021suc,Becattini:2021iol}, i.e., $\varpi_{\alpha\beta} \rightarrow (\partial_{\alpha} u_{\beta} - \partial_{\beta} u_{\alpha}) / (2T)$ and $\xi_{\alpha\beta} \rightarrow (\partial_{\sigma} u_{\alpha} + \partial_{\alpha} u_{\sigma}) / (2T)$. $n^{\mu}$ is chosen as the time unit vector $n^{\mu} = \delta_{0}^{\mu}$, and the mass $m$ is still chosen as the mass of $\Lambda$ hyperons \citep{Palermo:2024tza}. In the current work, we take $n^{\mu} \rightarrow u^{\mu}$ in this scenario for simplicity.
\end{itemize}
Note that the contributions from hadronic cascades are neglected in the above three scenarios. For the influence of hadronic scattering on spin polarization, we refer to Ref. \citep{Sung:2024vyc}. In the current work, we follow studies in relativistic nucleus-nucleus collisions and consider the above three scenarios.

The spin polarization rate $\vec{P}^{*}$ is measured in the rest frame of $\Lambda$ hyperons in experiments and can be obtained using the polarization pseudo-vector $\mathcal{S}^{\mu}$,
\begin{eqnarray}
\vec{P}^{*}(\mathbf{p}) & = & \vec{P}(\mathbf{p}) - \frac{\vec{P}(\mathbf{p}) \cdot \vec{p}}{p^{0}(p^{0} + m)}\vec{p}, \nonumber \\
P^{\mu}(\mathbf{p}) & \equiv & \frac{1}{\mathfrak{s}}\mathcal{S}^{\mu}(\mathbf{p}),
\end{eqnarray}
where $\mu=1,2,3$ represents the spin polarization rate along the $x$, $y$, and $z$ directions, respectively, and $\mathfrak{s}=1/2$ is the spin of $\Lambda$ hyperons. In relativistic nucleus-nucleus collisions, the spin of particles is globally polarized along the direction of the initial orbital angular momentum. However, in p+Pb collisions, the spin of particles is not naturally polarized along a certain direction collectively. Therefore, we will study the local spin polarization along the beam direction $P^{*z}$ in the current work. We define the azimuthal angle-dependent spin polarization along the beam direction as
\begin{eqnarray}
\langle P_{z}(\phi_{p}) \rangle & = & \frac{\int p_{T} dp_{T} d\eta [\Phi(\mathbf{p}) P^{*z}]}{\int p_{T} dp_{T} d\eta \Phi(\mathbf{p})}.\label{eq:Pz_phi}
\end{eqnarray}
The $p_{T}$ dependent and multiplicity dependent second Fourier sine coefficients of spin polarization along the beam direction can be defined as follows,
\begin{eqnarray}
\left\langle P_{z} \sin 2(\phi_{p} - \Psi_{2}) \right\rangle (p_{T}) & = & \frac{\int d\eta d\phi_{p} \widetilde{P^{z}}}{\int d\eta d\phi_{p} \Phi(\mathbf{p})}, \nonumber \\
\left\langle P_{z} \sin 2(\phi_{p} - \Psi_{2}) \right\rangle & = & \frac{\int p_{T} dp_{T} d\eta d\phi_{p} \widetilde{P^{z}}}{\int p_{T} dp_{T} d\eta d\phi_{p} \Phi(\mathbf{p})},\label{eq:Pz_sin}
\end{eqnarray}
where $\widetilde{P^{z}} \equiv \Phi(\mathbf{p}) P^{*z} \sin 2(\phi_{p} - \Psi_{2})$, 
and $\eta$ is the momentum pseudo-rapidity.

\begin{table}
\caption{The average number of charged particles in each multiplicity interval. $\langle N_{\text{ch}}\rangle_{\textrm{exp}}$ is the experimental data measured by the CMS collaboration \citep{chenyan2024talk}, while $\langle N_{\text{ch}}\rangle_{\text{CLVisc}}$ is computed by our model. \label{tab:The-average-number}}
\centering{}%
\begin{tabular}{c|c|c}
\hline 
$\;$Multiplicity intervals$\;$ & $\;\langle N_{\text{ch}}\rangle_{\textrm{exp}}\;$ & $\;\langle N_{\text{ch}}\rangle_{\text{CLVisc}}\;$\tabularnewline
\hline 
\hline 
{[}185,250) & 203.3 & 204.2\tabularnewline
\hline 
{[}150,185) & 163.6 & 164.5\tabularnewline
\hline 
{[}120,150) & 132.7 & 133.57\tabularnewline
\hline 
{[}60,120) & 86.7 & 87.7\tabularnewline
\hline 
{[}3,60) & 40 & 29.3\tabularnewline
\hline 
\end{tabular}
\end{table}

\begin{figure*}
\includegraphics[scale=0.5]{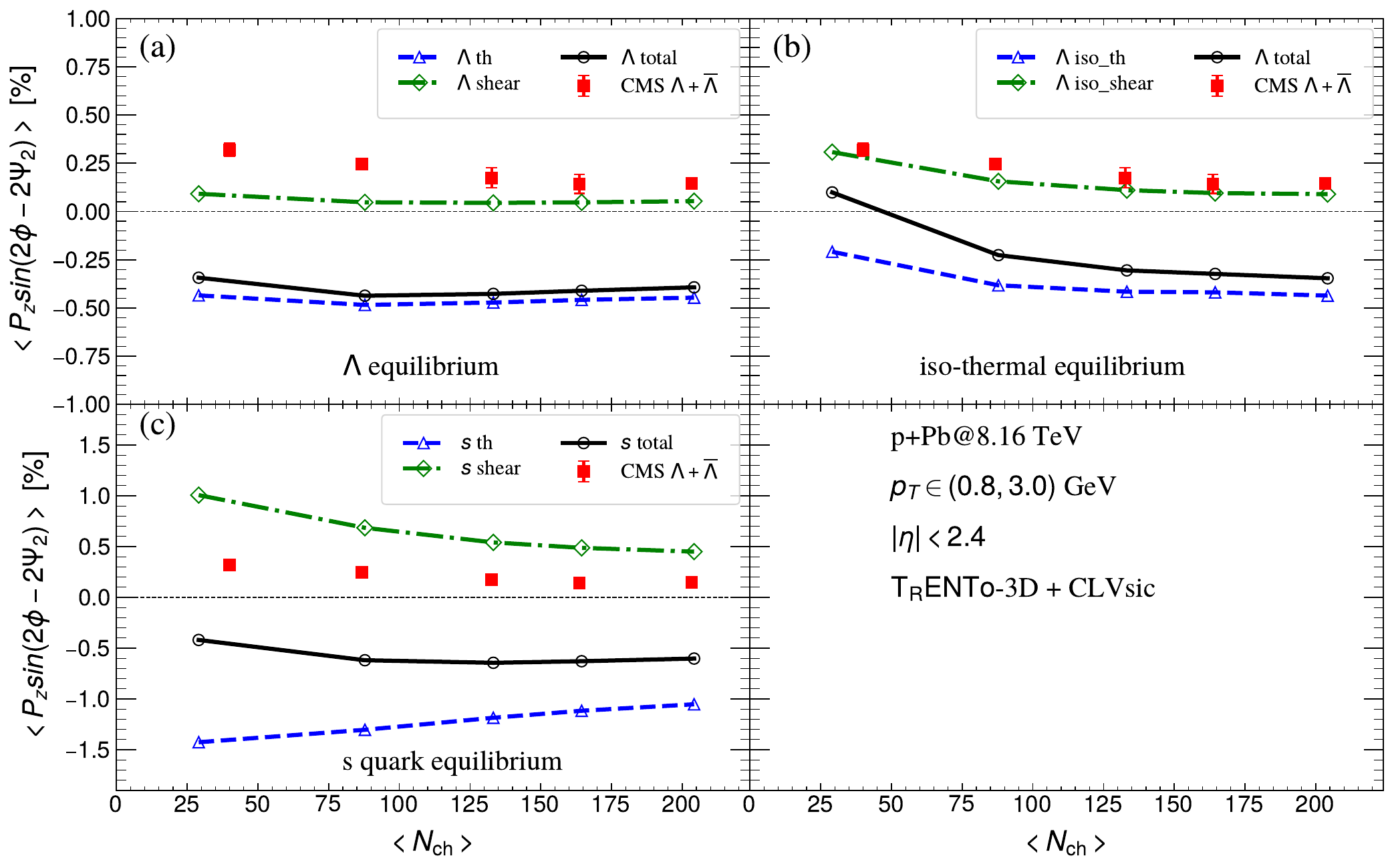}
\caption{The second Fourier sine coefficient of the spin polarization $\langle P_{z} \sin(2\phi_{p} - 2\Psi_{2}) \rangle$ for $\Lambda$ hyperons as a function of multiplicity in p+Pb collisions at $\sqrt{s_{NN}} = 8.16$ TeV in (a) $\Lambda$ equilibrium, (b) $s$ quark equilibrium, and (c) iso-thermal equilibrium scenarios. The blue triangular, green diamond-shaped, and black circular points denote the spin polarization induced by the thermal-vorticity tensor, thermal-shear tensor, and total effects, respectively. The red data points are given by CMS experiments \citep{chenyan2024talk}. The results are set up with $p_{T} \in (0.8, 3.0)$ GeV and pseudo-rapidity $|\eta| < 2.4$. \protect\label{fig:P2z_muliti}}
\end{figure*}

\begin{figure*}
\includegraphics[scale=0.5]{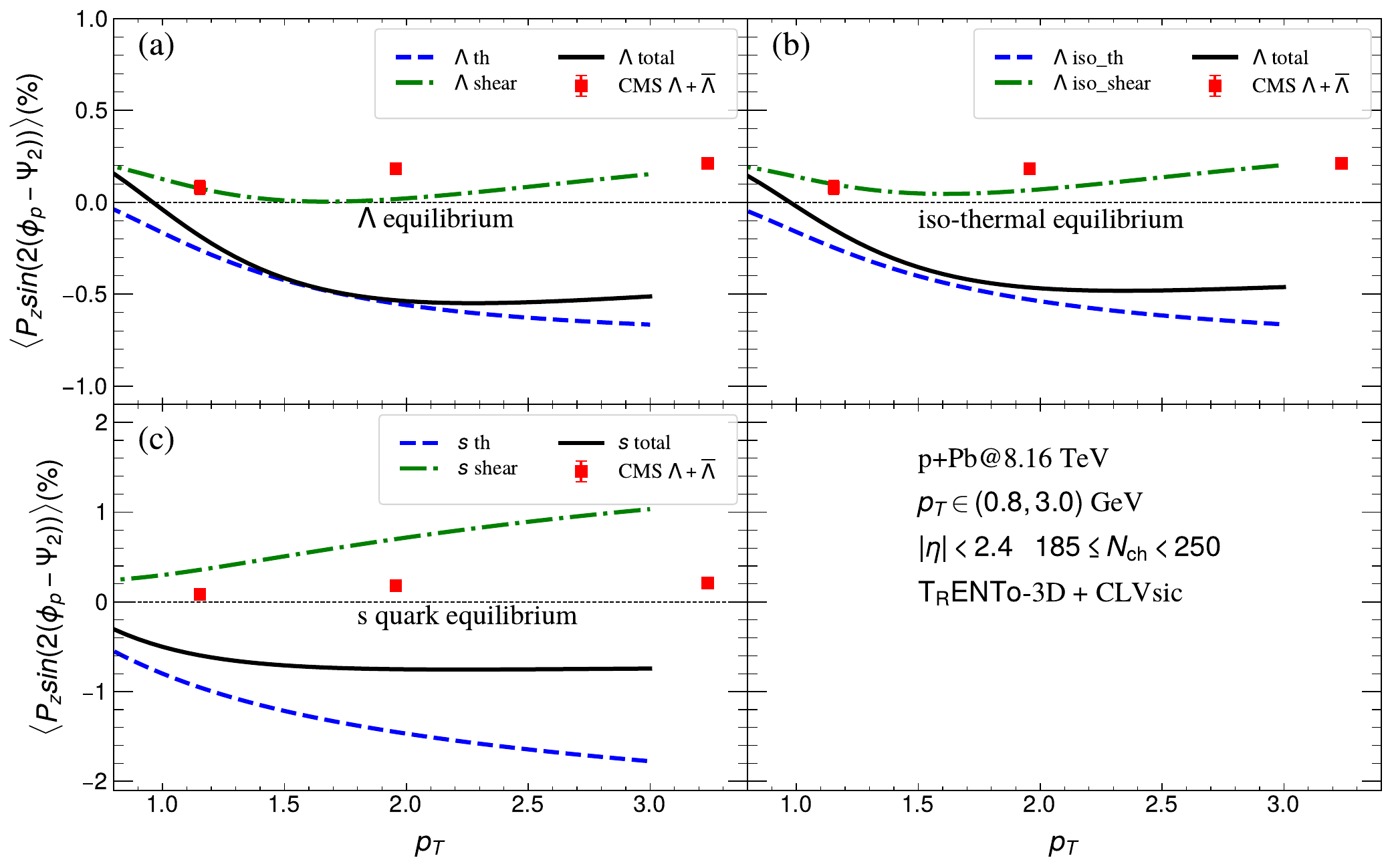}
\caption{The second Fourier sine coefficient of the spin polarization $\langle P_{z} \sin(2\phi_{p} - 2\Psi_{2}) \rangle$ for $\Lambda$ hyperons as a function of transverse momentum $p_{T}$ in p+Pb collisions at $\sqrt{s_{NN}} = 8.16$ TeV in (a) $\Lambda$ equilibrium, (b) $s$ quark equilibrium, and (c) iso-thermal equilibrium scenarios. The multiplicity is chosen as $185 \leq N_{\text{ch}} < 250$, with $p_{T} \in (0.8, 3.0)$ GeV and pseudo-rapidity $|\eta| < 2.4$. The blue dashed, green dotted-dashed, and black solid lines represent the spin polarization induced by the thermal-vorticity tensor, thermal-shear tensor, and total effects, respectively. The red data points are given by CMS experiments \citep{chenyan2024talk}. \label{fig:P2z_pt}}
\end{figure*}

\section{Numerical setup \label{sec:Numerical-framework}}

In this work, we implement the 3+1D CLVisc hydrodynamics model \citep{Pang:2012he,Wu:2021fjf,Wu:2022mkr} with the parameterized \trento-3D model \citep{Soeder:2023vdn,Moreland:2018gsh,Moreland:2014oya,Ke:2016jrd} as initial conditions to simulate the hydrodynamic evolution in p+Pb collisions at $\sqrt{s_{NN}} = 8.16$ TeV.

Given the collision system and energy, the \trento-3D model will sample the transverse position of nucleons from the Woods-Saxon distribution at mid-rapidity. Then the participants are sampled according to the inelastic nucleon-nucleon cross section, which is tuned to fit the Particle Data Group data. The thickness function of the two nuclei is obtained from the wounded participant nucleons,
\begin{eqnarray}
T_{A/B} & = & \sum_{i=1}^{N_{A/B}} \gamma_{i} T_{p}(\mathbf{x}_{\perp}, \mathbf{x}_{\perp}^{i}, \omega),
\end{eqnarray}
where $N_{A/B}$ is the number of wounded nucleons in the projectile (target) nucleus, $\gamma_{i}$ is a random number sampled from the unit $\Gamma$ distribution with variance $1/k$. The proton thickness function $T_{p}$ is a 2-dimensional unit Gaussian distribution with width $\omega$ centered at $\mathbf{x}_{\perp}$. If we consider the constituents of the nucleons, the thickness function of the projectile and target nuclei will be replaced by
\begin{eqnarray}
T_{A/B}(\mathbf{x}_{\perp}) & = & \sum_{i=1}^{N_{A/B}} \frac{1}{n_{c}} \sum_{q=1}^{n_{c}} \gamma_{q} \frac{e^{-(\mathbf{x}_{\perp} - \mathbf{x}_{\perp}^{i} - \mathbf{s}_{q})^{2}/2v^{2}}}{2\pi v^{2}},
\end{eqnarray}
where we choose $n_{c} = 3$ as the number of constituents in each nucleon and $v$ is the width of the constituent distribution. The transverse position of a constituent within nucleons $\mathbf{s}_{q}$ is sampled from a Gaussian distribution. The width of this distribution is chosen to ensure each nucleon exhibits a Gaussian distribution with width $\omega$.

The initial thermodynamic variables can be expressed as a function of entropy density $s$ through the given equations of state. The initial entropy density as a function of $\mathbf{x}_{\perp}$ and spatial rapidity $\eta_{s}$ is assumed to be,
\begin{eqnarray}
s(\mathbf{x}_{\perp}, \eta_{s})|_{\tau = \tau_{0}} & = & K s(\mathbf{x}_{\perp}) g(\mathbf{x}_{\perp}, y) \frac{dy}{d\eta_{s}}, \label{eq:Trento-3d}
\end{eqnarray}
where $\tau_{0}$ is the initial proper time, $K$ is tuned to fit the $dN_{\text{ch}}/d\eta$ spectrum of the final charged hadrons, $s(\mathbf{x}_{\perp}) \propto \left(\frac{T_{A}^{a} + T_{B}^{a}}{2}\right)^{1/a}$ with $a$ being a constant, and $g(\mathbf{x}_{\perp}, y)$ is a function of momentum rapidity $y$ with $g(\mathbf{x}_{\perp}, 0) = 1$. Here, $\frac{dy}{d\eta_{s}} \simeq J \cosh \eta / \sqrt{1 + J^{2} \sinh^{2} \eta}$ with $J \simeq \langle p_{T} \rangle / \langle m_{T} \rangle$ being the Jacobi factor. In the \trento-3D model, the function $g(\mathbf{x}_{\perp}, y)$ is evaluated by the inverse Fourier transformation of its cumulant-generating function $\tilde{g}(\mathbf{x}_{\perp}, k)$, i.e., $g(\mathbf{x}_{\perp}, y) = (2\pi)^{-1} \int dk \, e^{-iky} \tilde{g}(\mathbf{x}_{\perp}, k)$, and $\tilde{g}(\mathbf{x}_{\perp}, k)$ is parameterized as,
\begin{eqnarray}
\log \tilde{g}(\mathbf{x}_{\perp}, k) & = & i \mu k - \frac{1}{2} \sigma^{2} k^{2} - \frac{1}{6} i \gamma \sigma^{3} k^{3} \exp\left(-\frac{1}{2} \sigma^{2} k^{2}\right) \nonumber \\
&+& \ldots,
\end{eqnarray}
where $\mu, \sigma, \gamma$ are the shift, width, and skewness of the rapidity distribution, respectively and $\ldots$ denotes higher-order corrections. In the current study, we adopt the relative-skewness model \citep{Soeder:2023vdn, Ke:2016jrd} to generate the $g(\mathbf{x}_{\perp}, y)$ and set local rapidity distribution's shift $\mu_0 = 0$, width $\sigma_0 = 2.9$, skewness $\gamma_0 = 6$.

The subsequent evolution of the system is simulated by the 3+1D CLVisc hydrodynamics model. In the current work, we focus on the energy-momentum conservation equations $\partial_{\mu} T^{\mu\nu} = 0$, where $T^{\mu\nu}$ is the energy-momentum tensor and includes the bulk viscous pressure $\Pi$ and shear-stress tensor $\pi^{\mu\nu}$. The evolution of $\Pi$ and $\pi^{\mu\nu}$ is described by the second-order Israel-Stewart equations \citep{Denicol:2018wdp, Baier:2007ix}. We follow Ref. \citep{Bernhard:2016tnd} to adopt the temperature-dependent shear viscosity $\eta_{v}(T)$ and bulk viscosity $\zeta(T)$. The equations of state are provided by the HotQCD collaboration \citep{HotQCD:2014kol}.

As the system expands according to the hydrodynamic equations, the temperature decreases. At the late stage of the evolution, one can define the (chemical) freeze-out hypersurface, on which the local temperature reaches the freeze-out temperature $T_{f} = 154$ MeV. When fluid cells pass through this freeze-out hypersurface, they will form soft particles based on the Cooper-Frye formula \citep{McNelis:2021acu, McNelis:2019auj, Denicol:2018wdp}. After the particlization of the fluid, these thermal hadrons will produce final  hadrons through resonance decay. To simplify, we neglect the afterburner effects in the multiplicity calculation and include the  Simulating Many Accelerated Strongly-interacting Hadrons (SMASH) model for afterburner in the differential $p_T$ anisotropic flow calculation. Once we obtain the hydrodynamic variables at the freeze-out hypersurface, we can calculate the spin polarization of $\Lambda$ hyperons according to Eqs.~(\ref{eq:th_shear}).

We have run $10^{5}$ minimum bias events to divide the centrality. The centrality-dependent pseudo-rapidity distributions of charged hadrons and elliptic flow for $\Lambda$ hyperons computed by our model are consistent with the experimental measurements \citep{CMS:2017shj, ALICE:2018wma, CMS:2018loe} as shown in Fig.~\ref{fig:dNdeta}. Following the experimental measurement, we divide the multiplicity into 5 intervals: $3 \leq N_{\text{ch}} < 60$, $60 \leq N_{\text{ch}} < 120$, $120 \leq N_{\text{ch}} < 150$, $150 \leq N_{\text{ch}} < 185$, $185 \leq N_{\text{ch}} < 250$, where $N_{\text{ch}}$ is the number of charged particles in the momentum region $p_{T} > 0.4$ GeV and $|\eta| < 2.4$. There are at least $2000$ minimum bias events in each interval for calculating the spin polarization of $\Lambda$ hyperons. We also compared the average number of charged particles in each multiplicity interval computed by our simulation with the experimental data, as shown in Table \ref{tab:The-average-number}. The QGP may not be generated in low multiplicity in p+Pb collisions. Therefore, hydrodynamic descriptions may be inapplicable for this case. That is why the average value of multiplicity in the interval $3 \leq N_{\text{ch}} < 60$ computed by our simulation shows a discrepancy with $N_{\text{ch}}$ measured by the experiments.

We follow the experimental measurements and choose the integration region as $p_{T} \in [0.8, 3.0]$ GeV, $\eta \in [-2.4, 2.4]$, and $\phi \in [0, 2\pi]$ for computing $\langle P_{z}(\phi_{p}) \rangle$ and $\left\langle P_{z} \sin 2(\phi_{p} - \Psi_{2}) \right\rangle$ in Eqs.~(\ref{eq:Pz_phi}, \ref{eq:Pz_sin}).


\section{Numerical results \label{sec:Numerical-results}}

In this section, we present the spin polarization of $\Lambda$ hyperons along the beam direction as functions of multiplicity $N_{\text{ch}}$, transverse momentum $p_{T}$, azimuthal angle $\phi_{p}$ and  pseudo-rapidity $\eta$  in p+Pb collisions at $\sqrt{s_{NN}} = 8.16$ TeV, simulated by 3+1D CLVisc hydrodynamics model with \trento-3D initial conditions.

\begin{figure*}
\includegraphics[scale=0.35]{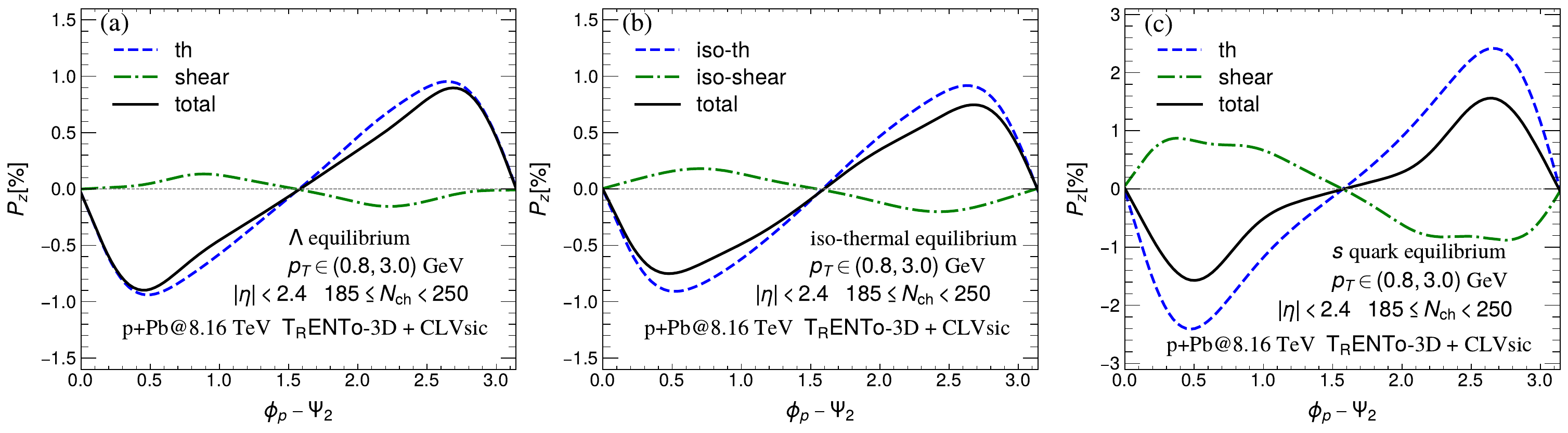}
\caption{The local spin polarization of $\Lambda$ hyperons along the beam direction, $P_{z}$, as a function of azimuthal angle $\phi_{p}-\Psi_2$ in p+Pb collisions at $\sqrt{s_{NN}} = 8.16$ TeV in three scenarios. The events are selected where $185 \le N_{\text{ch}} \le 250$, $p_{T} > 0.4$ GeV and $|\eta| < 2.4$. Following the same color assignments as in Fig.~\ref{fig:P2z_muliti}. \label{fig:P2z_phi}} 
\end{figure*}

In Fig.~\ref{fig:P2z_muliti}, we plot $\langle P_{z} \sin(2\phi_{p} - 2\Psi_{2}) \rangle$ for $\Lambda$ hyperons as a function of  multiplicity $N_{\text{ch}}$ in p+Pb collisions at $\sqrt{s_{NN}} = 8.16$ TeV in three scenarios. First, we observe that the shear-induced polarization in these three scenarios, denoted by the green dash-dotted lines in Fig.~\ref{fig:P2z_muliti}, always gives a positive contribution to $\langle P_{z} \sin(2\phi_{p} - 2\Psi_{2}) \rangle$. In contrast, the polarization induced by the thermal vorticity, denoted by the blue dashed lines in Fig.~\ref{fig:P2z_muliti}, provides an opposite contribution compared to the shear-induced polarization. These are similar to spin polarization in relativistic nucleus-nucleus collisions \citep{Fu:2021pok, Becattini:2021iol, Yi:2021ryh, Wu:2022mkr}. We then find that the total $\langle P_{z} \sin(2\phi_{p} - 2\Psi_{2}) \rangle$ computed by our model in three scenarios, which are denoted by the black solid lines in Fig.~\ref{fig:P2z_muliti}, is always negative except for the result in low multiplicity in the iso-thermal equilibrium scenario. The results in the three scenarios are inconsistent with the data from the LHC-CMS experiments. Our findings indicate that the spin polarization induced  by the thermal vorticity and shear tensors cannot explain the experimental data in p+Pb collisions well.

Moreover, we observe that the total $\langle P_{z} \sin(2\phi - 2\Psi_{2}) \rangle$ computed by our model in both $\Lambda$ and $s$ quark equilibrium scenarios is insensitive to $N_{\text{ch}}$. This is because the dependence of anisotropic expansion on multiplicity is much weaker in p+Pb systems compared to that in relativistic nucleus-nucleus systems. Since the particle's mass in Eqs.~(\ref{eq:th_shear}) is chosen as $m_{s} = 0.3$ GeV in the $s$ quark equilibrium scenario, the magnitude of spin polarization induced by both thermal vorticity and shear tensors in $s$ quark equilibrium scenario should be much larger than those in the $\Lambda$ equilibrium scenario. This agrees with the results shown in Fig.~\ref{fig:P2z_muliti}. On the other hand, the spin polarization induced by thermal vorticity and shear tensors in the $s$ quark equilibrium scenario does not simply scale up the results in the $\Lambda$ equilibrium scenario. 

Interestingly, we find that the total $\langle P_{z} \sin(2\phi - 2\Psi_{2}) \rangle$ decreases with growing $N_{\text{ch}}$ in the iso-thermal equilibrium scenario. Recalling that all effects related to $\nabla T$ are dropped in the iso-thermal equilibrium scenario, this observation indicates that the spin polarization induced by the terms proportional to $\nabla T$ is negative and decreases with increasing multiplicity. This is similar to the observation in relativistic nucleus-nucleus collisions \citep{Fu:2021pok, Yi:2021ryh}.

In Fig.~\ref{fig:P2z_pt}, we plot the $\langle P_{z} \sin(2\phi_{p} - 2\Psi_{2}) \rangle$ for $\Lambda$ hyperons as a function of transverse momentum $p_{T}$ in the high multiplicity region $185 \le N_{\text{ch}} \le 250$ in p+Pb collisions at $\sqrt{s_{NN}} = 8.16$ TeV. Similar to Fig.~\ref{fig:P2z_muliti}, the spin polarization induced by thermal vorticity is always negative, whereas the shear-induced polarization is always positive. In both $\Lambda$ and iso-thermal equilibrium scenarios, when $p_{T}$ increases, the shear-induced polarization decreases at low $p_{T}$ but increases in the high transverse momentum region $p_{T} \gtrsim 1.6$ GeV. As a consequence, the total $\langle P_{z} \sin(2\phi_{p} - 2\Psi_{2}) \rangle$ in these two scenarios changes from positive to negative at $p_{T} \approx 1.0$ GeV. Meanwhile, the shear-induced polarization in the $s$ quark scenario grows as $p_{T}$ increases. The total $\langle P_{z} \sin(2\phi_{p} - 2\Psi_{2}) \rangle$ in this scenario decreases with growing $p_{T}$ and becomes almost flat in the region $p_{T} \gtrsim 1.6$ GeV.

In Fig.~\ref{fig:P2z_phi}, we present the spin polarization along the beam direction $P_{z}$ contributed by different sources as a function of azimuthal angle $\phi_{p} - \Psi_{2}$ in the high multiplicity region, $185 \le N_{\text{ch}} \le 250$, in three different scenarios. The spin polarization induced by the thermal vorticity tensor consistently follows a $-\sin 2\phi_{p}$ form, while the shear-induced polarization follows a $\sin 2\phi_{p}$ form in the $\Lambda$ and iso-thermal equilibrium scenarios. We find that in the $s$ quark equilibrium scenario, spin polarization induced by thermal vorticity and shear tensors does not strictly follow the forms mentioned. Since the spin polarization induced by the thermal vorticity tensor dominates in the high multiplicity region, the total $P_{z}$ approximately follows a $-\sin 2\phi_{p}$ form in these three scenarios. Additionally, we find that $P_{z}$ induced by different sources in both $\Lambda$ and iso-thermal equilibrium scenarios agrees with each other. This is because the magnitude of $\nabla T$ in the freeze-out hypersurface in the high multiplicity region is much smaller than it in the low multiplicity region.
\begin{figure}
\includegraphics[scale=0.4]{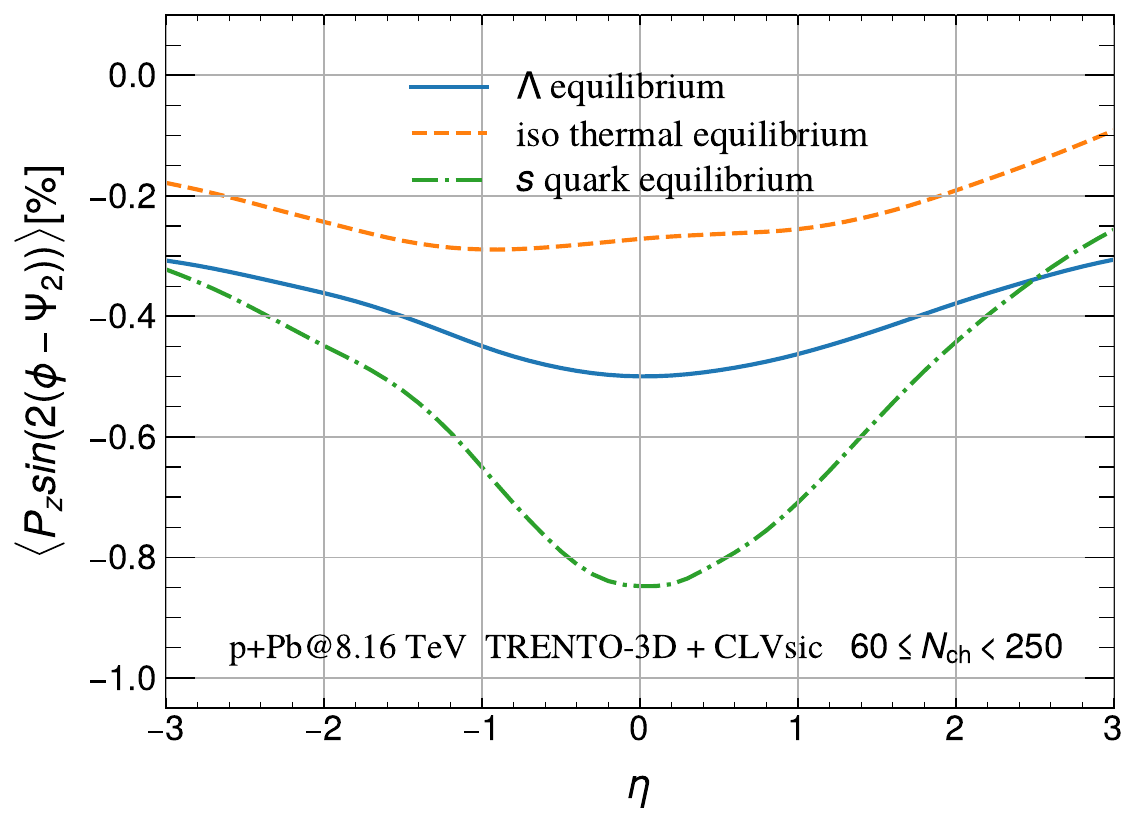}

\caption{The total second Fourier sine coefficient of the spin polarization
$\langle P_{z}\sin(2\phi_{p}-2\Psi_{2})\rangle$ for $\Lambda$ hyperons
as a function of pseudo-rapidity $\eta$ in p+Pb collisions at $\sqrt{s_{NN}}=8.16$
TeV. The events are selected where $60\le N_{\text{ch}}\le250$. The
blue solid, green dotted-dashed, and orange dashed lines represent
the results in $\Lambda$ equilibrium, $s$ quark equilibrium, and
iso-thermal equilibrium scenarios, respectively.}\label{fig:eta_dependence}
\end{figure}

In Fig.~\ref{fig:eta_dependence}, we predict the $\langle P_{z}\sin(2\phi_{p}-2\Psi_{2})\rangle$ as a function of pseudo-rapidity $\eta$ in three scenarios. 
We observe that $\langle P_{z}\sin(2\phi_{p}-2\Psi_{2})\rangle$ is not symmetric under the transformation $\eta \rightarrow -\eta$, which 
arises from the asymmetry in the size of the proton and nucleus in the initial states.

\section{Conclusion and discussion \label{sec:Summary}}

In this work, we have investigated the spin polarization of $\Lambda$ hyperons along the beam direction in p+Pb collisions at $\sqrt{s_{NN}} = 8.16$ TeV. We follow the theoretical framework in our previous works \citep{Yi:2021ryh, Wu:2022mkr, Yi:2023tgg} and implement the 3+1D CLVisc hydrodynamics model with \trento-3D initial conditions. We consider three scenarios: $\Lambda$ equilibrium, $s$ quark equilibrium, and iso-thermal equilibrium scenarios. These three scenarios are used in studies of spin polarization of hyperons in relativistic nucleus-nucleus collisions. We compute $\left\langle P_{z} \sin 2(\phi_{p} - \Psi_{2}) \right\rangle$ as functions of multiplicity $N_{\text{ch}}$, transverse momentum $p_{T}$, pseudo-rapidity $\eta$ and $P_{z}$ as a function of azimuthal angle $\phi_{p} - \Psi_{2}$ in the three scenarios.

Our conculsion is as follows. We observe that the polarization induced by the thermal vorticity tensor always gives an opposite contribution compared to the shear-induced polarization in p+Pb collisions, similar to relativistic nucleus-nucleus collisions. Although the shear-induced polarization always provides a trend consistent with the experimental data, the total spin polarization cannot be well explained by the current model.


Before ending this work, we would like to discuss other possible effects excluded in the current study that may contribute to the spin polarization along the beam direction. Let us concentrate on the possible corrections to the modified Cooper-Frye formula in Eq.~(\ref{eq:CS}). In principle, the electromagnetic fields and $\nabla (\mu/T)$, with $\mu$ being the (baryon) chemical potential, can also contribute to the polarization pseudo-vector $\mathcal{S}^\mu$, e.g. see our previous works \citep{Yi:2021unq, Yi:2021ryh, Wu:2022mkr, Yi:2023tgg} based on quantum kinetic theory. However, since electromagnetic fields decay rapidly, they cannot provide significant local polarization. For estimation of spin polarization induced by electromagnetic fields, see Refs.~\citep{Muller:2018ibh, Guo:2019joy, Buzzegoli:2022qrr, Xu:2022hql, Peng:2022cya}. The spin polarization induced by $\nabla (\mu/T)$, known as the spin Hall effect, can play a role in low-energy collisions but is negligible in the high-energy region, also see Refs.~\citep{Liu:2020dxg,Yi:2021ryh,Ryu:2021lnx,Fu:2022myl,Wu:2022mkr,Fu:2022oup}  for details. Therefore, the spin Hall effect may also be safely neglected in p+Pb collisions. Additionally, second-order effects in the gradient expansion may also play a role in spin polarization \citep{Sheng:2024pbw, Fang:2024preparation}.
On the other hand, recently, interaction corrections to the modified Cooper-Frye formula have drawn lots of attention, e.g. see Refs.~\citep{Fang:2022ttm, Fang:2023bbw, Lin:2024zik} and references therein. These effects need to be systematically studied in the future.

\begin{acknowledgments}
We would like to thank Zhenyu Chen and Chenyan Li for helpful discussion.
This work is supported in part by the National Key Research and Development
Program of China under Contract No. 2022YFA1605500, by the Chinese
Academy of Sciences (CAS) under Grants No. YSBR-088 and by National Natural Science Foundation of China (NSFC) under Grant Nos. 12075235,
12225503, 11890710, 11890711, 11935007, 12175122, 2021-867, 11221504,
11861131009 and 11890714.  X.-Y.W. is supported in part by the National Science Foundation (NSF) within the framework of the JETSCAPE collaboration under grant number OAC-2004571 and in part by the Natural Sciences and Engineering Research Council of Canada (NSERC). J.Z. is supported in part by China Scholarship Council (CSC) under Grant No. 202306770009.
\end{acknowledgments}

\bibliographystyle{h-physrev}
\bibliography{qkt-ref20230407}

\end{document}